\documentclass[reprint,nofootinbib,superscriptaddress,amsmath,amssymb,aps]{revtex4-1}
\usepackage{graphicx}
\usepackage{rotating}
\usepackage{dcolumn}
\usepackage{bm}
\usepackage{color}
\usepackage{mathptmx, textcomp}
\usepackage[latin1]{inputenc}
\usepackage{braket}
\usepackage[colorlinks,linkcolor=magenta,anchorcolor=cyan,citecolor=blue]{hyperref}
\usepackage[multiple]{footmisc}
\newcommand{\fig}[1]{Fig.\ref{#1}}
\def\be{\begin{equation}}
\def\ee{\end{equation}}
\def\ba{\begin{eqnarray}}
\def\ea{\end{eqnarray}}

\def\nn{\nonumber}
\def\lf{\left}
\def\rt{\right}

\newcommand{\eq}[1]{(\ref{#1})}

\def\nn{\nonumber}\def\lf{\left}\def\rt{\right}\def\q{\theta} \def\w{\omega}  \def\y {\psi}   \def\p {\pi} \def\a {\alpha}  \def\d {\delta} \def\f {\phi}  \def\h {\eta} \def\j {\varphi}  \def\l {\lambda}  \def\x {\xi}    \def\m {\mu}  \def\pd {\partial}\def\p {\pi} \def \inf {\infty}  \def \e { \varepsilon}
\def\Q{\Theta}      \def\S {\Sigma}  \def\F {\Phi}  \def\L {\Lambda}    \def\grad{\nabla}\def\.{\cdot}
\def\math {\mathcal}
\begin{document}

\title{New version of the gedanken experiments to test the weak cosmic censorship in charged dilaton-Lifshitz black holes}
\author{Jie Jiang}
\email{jiejiang@mail.bnu.edu.cn}
\affiliation{Department of Physics, Beijing Normal University, Beijing, 100875, China}
\author{Ming Zhang}
\email{mingzhang@jxnu.edu.cn (corresponding author)}
\affiliation{Department of Physics, Jiangxi Normal University, Nanchang 330022, China}
\date{\today}

\begin{abstract}
In this paper, based on the new version of the gedanken experiments proposed by Sorce and Wald, we examine the weak cosmic censorship in the perturbation process of accreting matter fields for the charged dilaton-Lifshitz black holes. In the investigation, we assume that the black hole is perturbed by some extra matter source satisfied the null energy condition and ultimately settle down to a static charged dilaton-Lifshitz black hole in the asymptotic future. Then, after applying the Noether charge method, we derive the first-order and second-order perturbation inequalities of the perturbation matter fields. As a result, we find that the nearly extremal charged dilaton-Lifshitz black hole cannot be destroyed under the second-order approximation of perturbation. This result implies that the weak cosmic censorship conjecture might be a general feature of the Einstein gravity, and it is independent of the asymptotic behaviors of the black holes.

\end{abstract}
\maketitle

\section{Introduction}

The general relativity predicts the existence of the black hole. There is a central singularity for most of the black holes. However, the singularity will make the spacetime ill-defined and destroy the law of causality. Therefore, Penrose proposed the weak cosmic censorship conjecture (WCCC) to ensure the predictability in the physical spacetime region \cite{RPenrose}. This conjecture states that all singularities caused by the gravitational collapsing body must be hidden inside an event horizon such that it does not affect the causality outside the black hole. It also means that the black holes cannot be destroyed by any physical process once it is formed if there is a singularity inside the event horizon.  To test this conjecture, Wald proposed a gedanken experiment to check whether the Kerr-Newman (KN) black hole can be destroyed by absorbing a test particle \cite{Wald94}. As a result, they found that the extremal KN black holes cannot be overspun or overcharged in this process under the first-order approximation. However, there are two drawbacks to this discussion, i.e., the initial black hole is extreme and it is only at the level of the first-order perturbation. For this story to be truly consistent, Hubeny extended the discussion to the second-order case in the nearly extremal KN black holes and showed that the nearly black hole can be destroyed in this case \cite{Hubeny} when the second-order effects are neglected. Their result attracted lots of researchers to extend it into various theories \cite{1,2,3,4,5,B1,B2,B3, B4, B5, B6,B7,B8,B9,B10,B11,B12,B13,B14,B16,B17}.

Recently, Sorce and Wald pointed out that if we consider the second-order correction, the spacetime cannot be easily treated as a background and we need to consider the full dynamical process of the spacetime and perturbation matter. Therefore, they proposed a new version of the gedanken experiments to overspin or overcharge the nearly KN black holes \cite{SW}. Based on the Noether charge method \cite{IW}, they considered the second-order corrections of the energy, angular momenta, and charge of the RN black hole, and derived a perturbation inequality at second-order approximation. Then, they concluded that these nearly extremal KN black holes cannot be destroyed after the second-order inequality are taken into consideration.

Most recently, the discussion of the new version has also been extended into some other stationary black holes \cite{Jiang:2020alh,Jiang1,Jiang2,Jiang3,An:2017phb,Ge:2017vun,Jiang:2019ige,WJ,Jiang:2019vww,Jiang:2019soz,He:2019mqy}. Although all of them showed the validity of the WCCC for nearly extremal black holes under the second-order approximation of perturbation, there is still a lack of the general proof of the WCCC. However, most of the researches only focus on the asymptotic flat spacetimes. We want to ask whether the WCCC is a general property for the Einstein gravity and it is independent of the asymptotic behaviors of the spacetime. Therefore, it is necessary for us to test the WCCC in the situation with different asymptotic behaviors. In this paper, we would like to consider the asymptotic Lifshitz black hole solution in Einstein-dilaton gravity and check whether the black hole can be destroyed by the new version of the gedanken experiments.

The remainder of this paper is organized as follows: in section \ref{sec2}, we review the spacetime geometry of the charged dilaton-Lifshitz black holes and discuss the perturbation from the physical process of accreting matter in this spacetime. In section \ref{sec3}, based on the Noether charge method as well as the null energy condition for all of the matter fields, we derive the first two order perturbation inequalities of the perturbation matter fields. In section \ref{sec4}, we discuss the possibility to destroy the nearly extremal black holes in the above physical process under the second-order approximation of the perturbation. Section \ref{sec5} is devoted to our conclusions.

\section{Linearly charged dilaton-Lifshitz black holes with the spherical perturbation}\label{sec2}

In this paper, we would like to test the WCCC of the asymptotic Lifshitz black holes by using the new version of the gedanken experiments proposed by Sorce and Wald \cite{SW}. In this section, we first review the four-dimensional charged dilaton-Lifshitz black hole solution in Einstein-dilaton gravity coupled to a linear Maxwell electrodynamics and two Lifshitz supporting gauge fields \cite{JS}. The Lagrangian four-form of this theory can be expressed as
\begin{equation}\label{PL}
	\bm{L}=\frac{\bm{\epsilon}}{16 \pi}\left(R-2\L-2\grad_a\F\grad^a\F-\sum_{i=1}^{3}e^{-2\a_i\F}\math{H}_i\right)+\bm{L}_\text{mt}\,,
\end{equation}
in which $\a_i$ is some coupling constant fixed as
\ba\begin{aligned}
\a_1=-\sqrt{z-1}\,,\ \ \ \ \a_2=\frac{2}{\sqrt{z-1}}\,,\ \ \ \a_3=\frac{1}{\sqrt{z-1}}
\end{aligned}\ea
with $z>1$, $\F$ is the dilaton field, $R$ is the Ricci scalar related to the metric $g_{ab}$, and $\bm{L}_\text{mt}$ is the Lagrangian of the extra matter source, $\math{H}_1=F_{ab}F^{ab}$, and $\math{H}_i=(H_i)_{ab}(H^i)^{ab}$ with $i=2,3$, where $\bm{F} = d \bm{A}$ is the strength of the electromagnetic field $\bm{A}$, and $\bm{H}_i=d\bm{B}_i$ with $i=2, 3$ can be regarded as the strength of the Lifshitz supporting gauge field $\bm{B}_i$. The equations of motion derived from the variation of above Lagrangian are given by
\ba\begin{aligned}\label{eoms}
&R_{ab}-\frac{1}{2}R g_{ab}-\L g_{ab}=8\p \left(T_{ab}^\text{EM}+T_{ab}^\text{B}+T_{ab}^\text{DIL}+T_{ab}^\text{mt}\right)\,,\\
&\grad_a G^{ba}_i=4\p j^b_i\,,\ \ \ \ { \grad^2\F+\frac{1}{2}\sum_{i=1}^{3}\a_i e^{-2\a_i\F}\math{H}_i=\y}
\end{aligned}\ea{
with
\ba\begin{aligned}
G_1^{ab}=e^{-2\a_1\F}F^{ab}\,,\quad G_{2,3}^{ab}=e^{-2\a_{2,3}\F}H^{ab}_{1,2}\,,
\end{aligned}\ea
}in which we have denoted the stress-energy tensors of the electromagnetic field, dilaton field, and supporting gauge fields as
\ba\begin{aligned}
T_{ab}^\text{EM}&=\frac{e^{-2\a_1\F}}{4\p}\left[F_{ac}F_b{}^c-\frac{1}{4}g_{ab}\math{H}_1\right]\,,\\
T_{ab}^\text{DIL}&=\frac{1}{4\p}\left(\grad_a\F\grad_b\F-\frac{1}{2}g_{ab}\grad_c\F\grad^c\F\right)\,,\\
T_{ab}^\text{B}&=\sum_{i=2}^{3}T_{ab}^i\,,
\end{aligned}\ea
with
\ba\begin{aligned}
T_{ab}^i&=\frac{e^{-2\a_i\F}}{4\p}\left[(H_i)_{ac}(H_i)_b{}^c-\frac{1}{4}g_{ab}\math{H}_i\right]\,,\\
\end{aligned}\ea
with $i=2,3$. Moreover, here $T_{ab}^\text{mt}$ is the stress-energy tensor of the accreting matter fields, $j_1^a$ and $j_{2,3}^a$ correspond to the current of the electromagnetic field and supporting gauge fields separately, and { $\y$} is the source of the dilaton field.

In the following, we consider the four-dimensional charged dilaton-Lifshitz solutions which are expressed as
\ba
\begin{aligned} \label{metric}
d s^{2} &=-\frac{r^{2z}}{L^{2z}}f(r) d v^2+\frac{2r^{z-1}}{L^{z-1}}
dv d r+r^{2}\left(d \theta^{2}+\sin ^{2} \theta d \j^{2}\right)\,, \\
\F(r)&=\sqrt{z-1}\ln\left(\frac{r}{b}\right)\,,\ \ \ \bm{A}=-\frac{q b^{2(z-1)}}{zr^z}dv\,,\\
\bm{B}_2&=\frac{q_2r^{z+2}}{(z+2)b^4}dv\,,\ \ \ \bm{B}_3=\frac{q_3r^z}{zb^2}dv
\end{aligned}
\ea
with the blackening factor
\begin{equation}\label{fr}	f(r)=1+\frac{L^2}{z^2r^2}-\frac{2M}{r^{z+2}}+\frac{q^2L^{2z}b^{2(z-1)}}{z r^{2(z+1)}}\,,
\end{equation}
and the constants
\ba\begin{aligned}
q_1^2=&\frac{b^4(z-1)(z+2)}{2L^{2z}}\,,\ \ \ \ \ q_3^2=\frac{b^2(z-1)}{L^{2(z-1)}z}\,,\\
&\quad\quad\L=-\frac{(z+1)(z+2)}{2L^2}\,.
\end{aligned}\ea
Here $b$ is some { positive} integral constant, $q$ and $M$ correspond to the electric charge and mass of the spacetime, separately. If there exists at least one root of the blackening factor $f(r)$, the solution describes a black hole. Otherwise, it describes a naked singularity. For the black hole case, the radius of the event horizon is the largest root of $f(r)$. If we also have $f'(r_h)=0$, the black hole becomes extreme and the conserved quantities have the constraints,
\ba\begin{aligned}
M&=\frac{b^2L^2r_h^{2z}+(z+1)b^2r_h^{2z}[L^2+z(z+2)r_h^2]}{(z+2)z^2b^2r_h^z}\,,\\
q^2&=\frac{r_h^{2z}[L^2+z(z+2)r_h^2]}{zL^{2z}b^{2(z-1)}}\,.
\end{aligned}\ea

In this paper, we would like to consider the situation when the static charged dilaton-Lifshitz black hole is perturbed by the spherically accreting matter fields which satisfy the null energy condition and it settles down to the dilaton-Lifshitz black hole with different parameters in the asymptotic future. A concrete example of this process is that a static black hole slowly accreting matter for a finite time and finally becomes another static black hole. In the following, we consider a family of above physical processes labeled by $\l$. Then, the dynamical fields $\f(\l)$ satisfies the equations of motion in \eq{eoms}. Here we denote $\f$ to the collection of $g_{ab}$, $\F$, $\bm{A}$, $\bm{B}_{2,3}$ as well as some extra matter fields. Generally, the spacetime in this physical process can be described by
\ba\begin{aligned}\label{ds2}
ds^2&=-\frac{r^{2z}}{L^{2z}}f(r,v,\l)dv^2+2\m(r,v,\l)dr dv\\
&+r^2(d\q^2+\sin^2\q d\j^2)\,,
\end{aligned}\ea
which satisfies
\ba\begin{aligned}
f(r,v,0)=f(r)\,,\ \ \ \ \m(r,v,0)=\frac{r^{z-1}}{L^{z-1}}
\end{aligned}\ea
for the background geometry. As mentioned above, at sufficiently late times, we assume that the black hole can also be described by the charged dilaton-Lifshitz solution with different parameters which can be labeled by $\l$, i.e., the line element can be expressed as
\ba
\begin{aligned}\label{S1dsa}
&f(r, v, \l)=f(r,\l)\\
&=1+\frac{L^2}{z^2r^2}-\frac{2M(\l)}{r^{z+2}}+\frac{q(\l)^2L^{2z}b(\l)^{2(z-1)}}{z r^{2(z+1)}}\,,\\
&\m(r,v,\l)=\frac{r^{z-1}}{L^{z-1}}
\end{aligned}\ea
at sufficiently late times. Then, the vector field $\x^a=(\pd/\pd v)^a$ becomes an Killing vector at late times. In this physical process, we also assume that the extra matter contains the sources of the dilaton field as well as the supporting gauge fields. Therefore, in the asymptotic future, the spacetime can be described by some different parameter $b(\l)$, { mass $M(\l)$ and charge $q(\l)$.} By virtue of this assumption, testing the weak cosmic censorship in this process is equivalent to checking whether the line element at late times describes a black hole geometry.

\section{Perturbation inequalities}\label{sec3}

In this section, we would like to derive some inequalities of the physical quantities for the perturbation at sufficiently late times under the second-order approximation. Different from the case of the asymptotic flat black holes, the mass of the black holes cannot be easily expressed like that in asymptotic spacetime. For simplification, we only consider the off-shell variation of the Einstein part. The Lagrangian four-form considered is given by
\ba\begin{aligned}\label{action}
\bm{L}=\frac{\bm{\epsilon}}{16\p}R\,.
\end{aligned}\ea
Following the notations in \cite{SW}, we will denote
\ba\begin{aligned}
\h=\h(0)\,,\ \ \d\h=\left.\frac{d\h}{d\l}\right|_{\l=0}\,,\ \ \d^2\h=\left.\frac{d^2\h}{d\l^2}\right|_{\l=0}
\end{aligned}\ea
for the physical quantity $\h(\l)$ in the family labeled by $\l$. The variation of above action gives
\ba\begin{aligned}\label{var1}
\d \bm{L}=\bm{E}_g^{ab} \d g_{ab}+d\bm{\Q}(g,\d g)\,,
\end{aligned}\ea
in which
\ba\begin{aligned}\label{EC}
\bm{E}_g^{ab}&=-\frac{1}{2}\bm{\epsilon}G^{ab}\,,\\
\bm{\Q}_{abc}^{}(g,\d g)&=\frac{1}{16\p}\bm{\epsilon}_{dabc}g^{de}g^{fg}\lf(\grad_g \d g_{ef}-\grad_e\d g_{fg}\rt)\,.
\end{aligned}\ea
Here $G_{ab}=R_{ab}-1/2R g_{ab}$ is the Einstein tensor. Using above expressions, the symplectic current three-form
\ba\begin{aligned}
\bm{\w}(g,\d_1 g,\d_2g)=\d_1\bm{\Q}(g,\d_2g)-\d_2\bm{\Q}(g,\d_1g)\,,
\end{aligned}\ea
can be expressed as
\ba\begin{aligned}\label{3w}
\bm{\w}_{abc}^{}&=\frac{1}{16\p}\bm{\epsilon}_{dabc}w^d
\end{aligned}\ea
with
\ba\begin{aligned}
w^a=P^{abcdef}\lf(\d_2g_{bc}\grad_d\d_1 g_{ef}-\d_1 g_{bc}\grad_d\d_2g_{ef}\rt)
\end{aligned}\ea
and
\ba\begin{aligned}
P^{abcdef}&=g^{ae}g^{fb}g^{cd}-\frac{1}{2}g^{ad}g^{be}g^{fc}\\
&-\frac{1}{2}g^{ab}g^{cd}g^{ef}-\frac{1}{2}g^{bc}g^{ae}g^{fd}+\frac{1}{2}g^{bc}g^{ad}g^{ef}\,.
\end{aligned}\ea

Using the Killing vector field $\x^a=(\pd/\pd v)^a$ of the background spactime, the Noether current three-form is defined as
\ba\begin{aligned}\label{J1}
\bm{J}_\x=\bm{\Q}(g,\math{L}_\x g)-\x\.\bm{L}\,.
\end{aligned}\ea
According to the calculation in \cite{Wald94}, it can be also written as
\ba\begin{aligned}\label{J2}
\bm{J}_\x=\bm{C}_\x+d\bm{Q}_\x\,,
\end{aligned}\ea
in which $C_\x=\x\.\bm{C}$ with
\ba\begin{aligned}\label{CTJ}
\bm{C}_{dabc}&=\bm{\epsilon}_{eabc}G_d{}^e\,,\\
\lf(\bm{Q}_\x^{}\rt)_{ab}&=-\frac{1}{16\p}\bm{\epsilon}_{abcd}\grad^c\x^d\,.
\end{aligned}\ea
Based on above results, the first two order variational identities can be calculated and they are expressed as
\ba\begin{aligned}\label{varid}
d[\d\bm{Q}_\x-\x\.\bm{\Q}(g,\d g)]&+\x\.\bm{E}_g^{ab}\d g_{ab}+\d \bm{C}_\x=0\,,\\
d[\d^2\bm{Q}_\x-\x\.\d\bm{\Q}(g,\d g)]&=\bm{\w}\lf(g,\d g,\math{L}_\x\d g\rt)\\
&-\d[\x\.\bm{E}_g^{ab}\d g_{ab}]-\d^2 \bm{C}_\x\,,
\end{aligned}\ea
in which we used the fact that $\x^a$ is the Killing vector of the background geometry.

\begin{figure}
\centering
\includegraphics[width=0.48\textwidth]{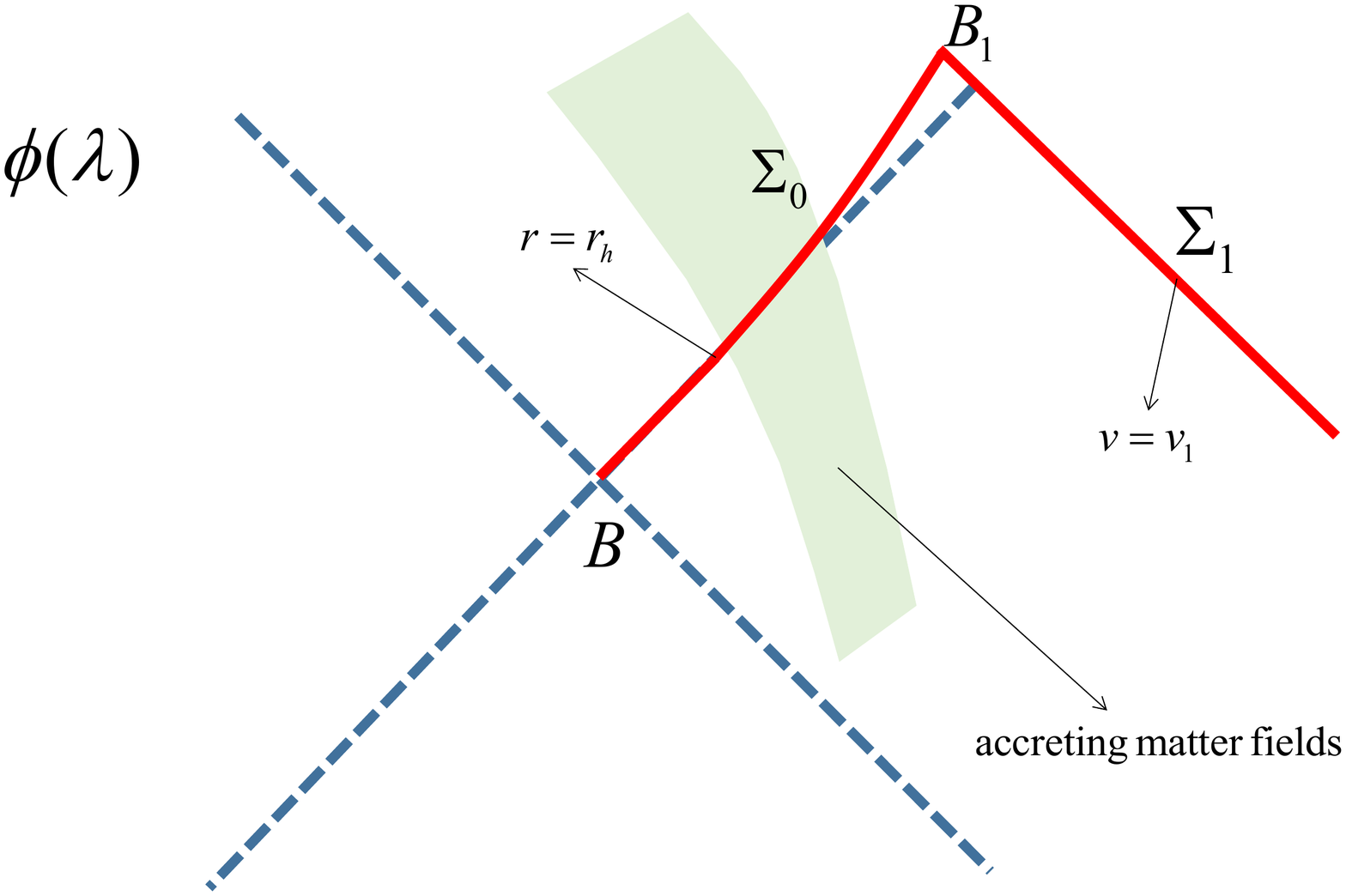}
\caption{{ Plot showing a dynamical configuration $\f(\l)$ where the static nonextremal black hole is perturbed by the spherically accreting matter fields. $\S_0$ is a hypersurface determined by $r=r_h$, where $r_h$ is the horizon radius of the background geometry $\f(0)$. Different from the hypersurface $\math{H}$ in \cite{SW}, $\S_0$ is not a null hypersurface in the configuration $\f(\l)$ with the line element $ds^2(\l)$ because $r_h$ is only the horizon radius of the background geometry.}}\label{fig}
\end{figure}

Since we assume that the process of accreting matter fields is in a finite time. We can choose a hypersurface which is made up of two portions $\S_0$ and $\S_1$ { (as shown in \fig{fig}),} where $\S_0$ starts at the cross-section $B$ where the perturbation vanishes and goes through the event horizon of the background spacetime (i.e., the hypersurface determined by $r=r_h=r_h(\l=0)$) to another cross-section $B_1$ at asymptotic future, and $\S_1$ is time-slice ($v=$constant) connected $B_1$ and spatial infinity $S_\inf$ at sufficiently late times.

Then, integration of the variational identities in \eq{varid} gives
\ba\begin{aligned}\label{var11}
&\int_{S_\inf}\lf[\d\bm{Q}_\x-\x\.\bm{\Q}(g,\d g)\rt]+\int_{\S_1}\x\.\bm{E}_g^{ab}\d g_{ab}\\
&+\int_{\S_1}\d \bm{C}_\x+\int_{\S_0}\d \bm{C}_\x=0
\end{aligned}\ea
and
\ba\begin{aligned}\label{var22}
&\int_{S_\inf}\left[\d^2 \bm{Q}_\x-\x\.\d\bm{\Q}(g,\d g)\right]+\int_{\S_1}\d[\x\.\bm{E}_g^{ab}\d g_{ab}]\\
&+\int_{\S_1}\d^2\bm{C}_\x+\int_{\S_0}\d^2\bm{C}_\x-\math{E}_{\S_1}-\math{E}_{\S_0}=0\,,
\end{aligned}\ea
where we denote
\ba\begin{aligned}
\math{E}_{\S_i}=\int_{\S_i}\bm{\w}(g,\d g,\math{L}_\x\d g)\,
\end{aligned}\ea
with $i=0,1$.

First of all, we turn to calculate the first equation \eq{var11}. Since the spacetime can be described by the line element in \eq{S1dsa} at sufficiently late times, we can explicitly calculate the first three terms in \eq{var11}. With straight calculation, the first term becomes
\ba\begin{aligned}\label{dM1}
\int_{S_\inf}\lf[\d\bm{Q}_\x^{}-\x\.\bm{\Q}^{}(g,\d g)\rt]=\frac{\d M}{L^{z+1}}\,.
\end{aligned}\ea
Also, we have
\ba\begin{aligned}\label{Tdg}
&T^{ab}(\l)\frac{d g_{ab}(\l)}{d\l}=-\frac{4(z-1)M'(\l)}{L^{z+2}r}\\
&+\frac{4(z-1)q(\l)b(\l)^{2z-3}}{zL^{2-2z}r^{2(z+1)}}\left[b(\l)q'(\l)+(z-1)q(\l)b'(\l)\right]
\end{aligned}\ea
on $\S_1$. Then, the second term can be further obtained and it is expressed as
\ba\begin{aligned}
\int_{\S_1}\x\.\bm{E}_g^{ab}\d g_{ab}&=\frac{(z-1)\ln r_h \d M}{r^{z+1}_h}\\
&+\frac{(z-1)q b^{2z-3}L^{z-1}}{z^2 r^z_h}[b\d q+(z-1)q \d b]\,.
\end{aligned}\ea
Using Eq. \eq{CTJ}, we have{
\ba\begin{aligned}\label{CL}
&\int_{\S_1} \bm{C}_\x(\l)=-\frac{(2z-1)q(\l)^2b(\l)^{2(z-1)}L^{z-1}}{2z^2r^z_h}\\
&-\frac{r_h^z}{2z^3L^{z+1}}\left(\frac{3z^3r_h}{z+2}+L^2(z^2-1)\right)-\frac{(z-1)M(\l)\ln r_h}{L^{z+1}}
\end{aligned}\ea
Using the above result, the third term can be obtained as
\ba\begin{aligned}
&\int_{\S_1} \d\bm{C}_\x=-\frac{(z-1)\ln r_h \d M}{r^{z+1}_h}\\
&-\frac{(2z-1)qb^{2z-3}L^{z-1}}{z^2r_h^z}[b\d q+(z-1)q\d b]\,.
\end{aligned}\ea
}Summing these results, the first-order variational identity becomes
\ba\begin{aligned}\label{ineq0}
&\frac{\d M}{L^{z+1}}-\frac{qb^{2z-3}L^{z-1}}{zr_h^z}[b\d q+(z-1)q\d b]=-\int_{\S_0}\d \bm{C}_\x\\
&= \d\left[\int_{\S_0}\bm{\tilde{\epsilon}} G_{ab}(dr)^a\x^b\right]=\d\left[\int_{\S_0}T_{ab}k^ak^bdv \hat{\bm{\epsilon}}\right]\,,
\end{aligned}\ea
where $T_{ab}$ is the stress energy tensor for all of the matter fields (including the dilaton field, electromagnetic field, supporting gauge fields, and extra matter fields), $k^a$ is a null vector field on $\S_0$ which is defined by
\ba\begin{aligned}
k(\l)=\left(\frac{\pd}{\pd v}\right)^a+\frac{r_h^{2z}f (r_h,v, \l)}{2L^{2z}\m(r_h, v, \l)}{ \left(\frac{\pd}{\pd r}\right)^a}\,,
\end{aligned}\ea
and the volume elements $\hat{\bm{\epsilon}}$ and $\tilde{\bm{\epsilon}}$ are defined by
\ba\begin{aligned}
\hat{\bm{\epsilon}}&=r^2\sin\q d\q\wedge d\j\,,\ \ \ \ \tilde{\bm{\epsilon}}=dv \wedge \hat{\bm{\epsilon}}\,.
\end{aligned}\ea

As mentioned in the last section, we assume that all of the matter fields satisfy the null energy condition, i.e., $T_{ab}(\l)k^a(\l)k^b(\l)\geq 0 $. Under the first-order approximation of perturbation, it gives $\d[T_{ab}k^ak^b]\geq 0$. Then, { the first-order variational identity \eq{ineq0} reduces to}
\ba\begin{aligned}\label{ineq1}
\d M-\frac{qb^{2z-3}L^{2z}}{zr_h^z}[b\d q+(z-1)q\d b]\geq0\,.
\end{aligned}\ea

{ The main purpose of this paper is to test whether the above perturbation process can destroy a nearly extremal black hole. When the first-order perturbation inequality is satisfied, in the next section, we will show that the WCCC cannot be violated under the first-order approximation. However, if the perturbation satisfies the optimal condition which saturates the first-order perturbation inequality \eq{ineq1}, i.e.,
\ba\begin{aligned}\label{exineq}
\d M-\frac{qb^{2z-3}L^{2z}}{zr_h^z}[b\d q+(z-1)q\d b]=0\,,
\end{aligned}\ea
the WCCC cannot be examined only considering the first-order approximation and the second-order approximation should be taken into account. Therefore, in the following, we would like to derive the second-order perturbation inequality under the first-order optimal condition. From the above discussions, the optimal condition also implies $\d\left[\sqrt{-g}T_{ab}(dr)^a\x^b\right]= 0$ on $\S_0$ by virtue of the null energy condition.} With a straightforward calculation, this gives $\pd_v \d f(r_h,v)=0$, in which we have defined the notation
\ba\begin{aligned}
\d \h(r,v)=\left.\frac{\pd \h(r,v,\l)}{\pd\l}\right|_{\l=0}\,.
\end{aligned}\ea
for the scalar function $\h(r,v,\l)$.

Next, we turn to evaluate the second-order variational identity in \eq{var22}. According to our assumption that the spacetime is static at sufficiently late times, the fifth term of \eq{var22} vanishes. For the first term, the straight calculation gives
\ba\begin{aligned}\label{dM2}
\int_{S_\inf}\left[\d^2 \bm{Q}_\x^{}-\x\.\d\bm{\Q}^{}(g,\d g)\right]=\frac{\d^2M}{L^{z+1}}\,.
\end{aligned}\ea
According to the results in Eqs. \eq{Tdg} and \eq{CL}, the second and third terms reduce to
\ba\begin{aligned}
&\int_{\S_1}\d[\x\.\bm{E}_g^{ab}\d g_{ab}]+\int_{\S_1}\d^2\bm{C}_\x\\
&=-\frac{qb^{2z-3}L^{z-1}}{zr_h^z}[b\d^2 q+(z-1)q\d^2 b]-\frac{b^{2(z-2)}L^{z-1}}{zr^z_h}\\
&\times[(2z-3)(z-1)q^2\d b^2+4(z-1)b q\d q \d b+b^2 \d q^2]\,,
\end{aligned}\ea
Considering the optimal condition of the first-order perturbation inequality, using the explicit expression of the metric \eq{ds2}, it is easy to verify that $\math{E}_{\S_0}=0$. Summing the above results, the second-order variational identity can be shown as
\ba\begin{aligned}\label{secgd}
&\frac{\d^2M}{L^{z+1}}-\frac{qb^{2z-3}L^{z-1}}{zr_h^z}[b\d^2 q+(z-1)q\d^2 b]-\frac{b^{2(z-2)}L^{z-1}}{zr^z_h}\\
&\times[(2z-3)(z-1)q^2\d b^2+4(z-1)b q\d q \d b+b^2 \d q^2]\\
&=-\int_{\S_0}\d^2\bm{C}_\x=\d^2\left[\int_{\S_0}\bm{\tilde{\epsilon}} T_{ab}(dr)^a\x^b\right]\\
&=\d^2\left[\int_{\S_0}T_{ab}k^ak^bdv \hat{\bm{\epsilon}}\right]\,.
\end{aligned}\ea
Because of the optimal condition of the first-order perturbation inequality, the null energy condition becomes $\d^2 [T_{ab}k^ak^b]\geq 0$ under the second-order approximation of perturbation. Then, the second-order variational identity reduces to
\ba\begin{aligned}\label{ineq2}
&\d^2M-\frac{qb^{2z-3}L^{2z}}{zr_h^z}[b\d^2 q+(z-1)q\d^2 b]-\frac{b^{2(z-2)}L^{2z}}{zr^z_h}\\
&\times[(2z-3)(z-1)q^2\d b^2+4(z-1)b q\d q \d b+b^2 \d q^2]\geq 0\,.
\end{aligned}\ea

\section{Gedanken experiments to destroy the nearly extremal black holes}\label{sec4}
Now we shall discuss the possibility to destroy the nearly extremal charged dialton-Lifshitz black holes in the physical process introduced in the previous sections. Because we assume that the spacetime settles down to a static state in the asymptotic future, checking the validity of the WCCC is equivalent to see whether the line element at sufficient late times also describes a black hole, i.e., there exists at least one root of the blacking factor $f(r,\l)$. To make it computable, we define a function
\ba
h(\l)=f(r_m(\l),\l)
\ea
to describe the minimal value of the blackening factor in the asymptotic future. Here $r_m(\l)$ is the minimal radius of the blackeing factor, and it can be obtained by
\ba\begin{aligned}\label{fprm}
{ \pd_rf(r_m(\l),\l)=0\,.}
\end{aligned}\ea
{ Using the explicit expression of blackening factor in Eq. \eq{S1dsa}, the above identity becomes
\ba\begin{aligned}\label{Mrm}
M(\l)=\frac{z(z+1)q^2(\l)b^{2z}(\l)L^{2z}+b^2(\l)L^2r_m^{2z}(\l)}{z^2(z+2)b^2(\l)r_m^z(\l)}\,.
\end{aligned}\ea
Under the zero-order approximation of $\l$, we have
\ba\begin{aligned}\label{Mqb}
M&=\frac{z(z+1)q^2b^{2z}L^{2z}+b^2L^2r_m^{2z}}{z^2(z+2)b^2r_m^z}\,.
\end{aligned}\ea
Taking the first-order variation of Eq. \eq{Mrm}, we can further obtain
\ba\begin{aligned}
\d M&=\frac{2b^{2z-3}L^{2z}(z+1)q[b\d q+(z-1)q\d b]}{z(z+2)r_m^z}\\
&+\frac{(b^3L^2r_m^{2z}-q^2z(z+1)b^{2z+1}L^{2z})\d r_m}{z(z+2)r_m^{z+1}b^3}\,,
\end{aligned}\ea
which implies
\ba\begin{aligned}
\d r_m&=\frac{z(z+2)r_m^{z+1}b^3\d M}{b^3L^2r_m^{2z}-q^2z(z+1)b^{2z+1}L^{2z}}\\
&-\frac{2q(z+1)L^{2z}r_m(b\d q+(z-1)q\d b)]}{b^{3-2z}L^2r_m^{2z}-q^2z(z+1)bL^{2z}}\,.
\end{aligned}\ea
}
Under the second-order approximation of perturbation, the minimal value of the blackening factor at late times can be expressed as
\ba\begin{aligned}\label{frm}
&h(\l)\simeq 1+\frac{L^2}{z(z+2)r_m^2}-\frac{L^2q^2}{(z+2)b^{2(z-1)}r_m^{2(z+1)}}\\
&-\frac{2\l}{r_m^{z+2}}\left(\d M-\frac{qb^{2z-3}L^{2z}}{zr_m^z}[b\d q+(z-1)q\d b]\right)\\
&-\frac{\l^2}{r_m^{z+2}}\left(\d^2 M-\frac{qb^{2z-3}L^{2z}}{zr_m^z}[b\d^2 q+(z-1)q\d^2 b]\right)\\
&+\frac{\l^2L^{2z}b^{2(z-2)}}{2zr_m^{2(z+1)}}[b^2\d q^2+(z-1)(2z-3)q^2\d b^2]\\
&+\frac{2\l^2(z-1)b^{2z-3}L^{2z}\d q\d b}{zr_m^{2(z+1)}}\,,
\end{aligned}\ea
{ where we have used Eq.\eq{Mqb} to replace $M$ by $r_m, q$ and $b$.} Because the physical process is only a perturbation of the background spacetime, the physical quantities at late times are only the small correction. Therefore, in order to destroy the black hole, the initial state must be a nearly extremal black hole. In the following, we consider the situation when the background spacetime is a nearly extremal black hole. Then, the position of the minimal value can be expressed as $r_m=(1-\epsilon)r_h$. With a similar setup as \cite{SW}, we assume that the parameter $\epsilon$ is agree with the first-order approximation of perturbation. { Then, we have
\ba\begin{aligned}\label{frm2}
f(r_m)&=f((1-\e)r_h)\\
&\simeq-\e r_h f'(r_h)+\frac{\e^2r_h^2}{2}f''(r_h)\\
&\simeq-\e r_h f'(r_m)-\e^2 r_h^2 f''(r_m)+\frac{\e^2r_h^2}{2}f''(r_h)\\
&=-\frac{\e^2r_h^2}{2}f''(r_h)\simeq -\frac{\e^2r_h^2}{2}f''(r_m)
\end{aligned}\ea
under the second-order approximation of $\epsilon$, i.e., we have neglected the higher-order term $O(\e^3)$ of $\e$. Using the explicit expression of the blackening factor in Eq. \eq{fr}, the left-hand side of the above equation gives
\ba\begin{aligned}
f(r_m)&=1+\frac{L^2}{z^2r^2_m}-\frac{2M}{r^{z+2}_m}+\frac{q^2L^{2z}b^{2(z-1)}}{z r^{2(z+1)}_m}\\
&=1+\frac{L^2}{z(z+2)r_m^2}-\frac{L^2q^2}{(z+2)b^{2(z-1)}r_m^{2(z+1)}}\,,
\end{aligned}\ea
where we have used Eq.\eq{Mqb} to replace $M$ by $r_m, q$ and $b$. For the right-hand side, we have
\ba\begin{aligned}
f''(r_m)&=-\frac{2(z+2)(z+3)M}{r^{z+4}}+\frac{6L^2}{z^2r_m^4}\\
&+2(1+z)(3+2z)z^{-1}b^{2z-2}L^{2z}q^2r_m^{-2(z+2)}\\
&=-\frac{2L^2}{zr_m^4}+\frac{2(z+1)L^{2z}q^2}{b^{2(1-z)}r_m^{2(z+2)}}\,.
\end{aligned}\ea
Summing the above results, Eq. \eq{frm2} becomes
\ba\begin{aligned}
&1+\frac{L^2}{z(z+2)r_m^2}-\frac{L^2q^2}{(z+2)b^{2(z-1)}r_m^{2(z+1)}}\\
&=\e^2r_h^2\left(\frac{L^2}{zr_m^4}-\frac{(z+1)L^{2z}q^2}{b^{2(1-z)}r_m^{2(z+2)}}\right)\\
&\simeq\frac{L^2\epsilon^2}{zr_h^2}-\frac{(z+1)\epsilon^2L^{2z}q^2}{b^{2(1-z)}r_h^{2(z+1)}}\,,
\end{aligned}\ea
under the second-order approximation of $\e$. In the last step, we have replaced $r_m=(1-\e) r_h$ by $r_h$ and neglected the higher-order term $O(\e^3)$. Using the above results, under the first-order approximation of perturbation, we have
\ba\begin{aligned}
h(\l)\simeq-\frac{2\l}{r_m^{z+2}}\left(\d M-\frac{qb^{2z-3}L^{2z}}{zr_m^z}[b\d q+(z-1)q\d b]\right)\,,
\end{aligned}\nn\\\ea
where we have neglected the higher-order terms $O(\e^2)$,  $O(\l^2)$ and $O(\l\e)$. Considering the first-order perturbation inequality \eq{ineq1}, we can see that $h(\l)\leq 0$ under the first-order approximation of perturbation. In the perturbation process, the signature of $h(\l)$ is totally determined by its leading order. If $h(\l)< 0$ under the first-order approximation, the signature is determined by the first-order approximation and the nearly extremal black holes cannot be destroyed in a perturbation process. However, under the optimal condition of the first-order perturbation inequality, we have $h(\l)=0$ under the first-order approximation. In this situation, the signature of $h(\l)$ cannot be determined by the first-order approximation and therefore we need to consider the second-order approximation of $h(\l)$. Combining the second-order perturbation inequality \eq{ineq2}, in the first-order optimal condition}, we can easily obtain
\ba\begin{aligned}
h(\l)&\leq -\frac{[b^3r_h^{2z}z\epsilon[L^2+(z+1)(z+2)r_h^2] -\l z^2 b^3 r_h^z \d M]^2}{b^6z^2r_h^{4z+2}[L^2+(z+1)(z+2)r_h^2]}\\
&\leq 0
\end{aligned}\ea
under the second-order approximation of perturbation. In the above calculation, we have replaced $r_m=(1-\e)r_h$ by $r_h$ because their difference only contributes some higher-order corrections. We can see that $h(\l)\leq 0$ under the second-order approximation. This result implies that the charged dilaton-Lifshitz black holes cannot be destroyed in the above physical process as long as the matter field satisfies the null energy condition.

\section{Conclusion}\label{sec5}
There are a lot of investigations to test the weak cosmic censorship in various spacetimes background based on the new version of the gedanken experiments. All of them showed the validity of the WCCC. However, there is still a lack of general proof of the WCCC even in general relativity. It is natural for us to ask whether its validity is independent of the asymptotic behaviors of spacetime. Therefore, in this paper, we examined the WCCC in the situation when the charged dilaton-Lifshitz black holes are perturbed by the spherically accreting matter which satisfies the null energy condition and finally settles to down to a static state at asymptotic future. Based on the Noether charge method, we first derived the first-order and second-order perturbation inequalities. As a result, we found that the charged dilaton-Lifshitz black holes cannot be destroyed by the above physical process under the second-order approximation of perturbation. Our result implies that the WCCC might be a general feature of the general relativity, and its validity does not depend on the asymptotic behavior of the black hole.

\section*{Acknowledgement}
This research was supported by National Natural Science Foundation of China (NSFC) with Grants No. 11775022 and 11873044.

\end{document}